\documentclass[aps,prl,reprint,groupedaddress]{revtex4-1}
\usepackage{graphicx}
\usepackage{dcolumn}
\usepackage{bm}
\usepackage{amsmath}
\usepackage{xfrac}
\usepackage{epstopdf}
\begin{document}


\title{Memory Recall and Spike Frequency Adaptation}


\author{James P. Roach}
\email[]{roachjp@umich.edu}
\affiliation{Neuroscience Graduate Program, University of Michigan}

\author{Leonard M. Sander}
\affiliation{Department of Physics, University of Michigan}
\affiliation{Center for the Study of Complex Systems, University of Michigan}

\author{Michal R. Zochowski}
\affiliation{Department of Physics, University of Michigan}
\affiliation{Center for the Study of Complex Systems, University of Michigan}
\affiliation{Biophysics Program, University of Michigan}

\date{\today}

\begin{abstract}
The brain can  reproduce memories from partial data; this ability is critical for memory recall.  The process of memory recall has been studied using auto-associative networks such as the Hopfield model. This kind of model reliably converges to stored patterns which contain the memory. However,  it is unclear how the behavior is controlled by the brain  so that after convergence to one configuration, it can  proceed with recognition of another one. In the Hopfield model this happens only through unrealistic changes of an effective global temperature that destabilizes all stored configurations. 
Here we show that spike frequency adaptation (SFA), a common mechanism affecting neuron activation in the brain, can provide  state dependent control of pattern retrieval. We demonstrate this in a Hopfield network modified to include  SFA, and also in a model network of biophysical neurons. In both cases SFA allows for selective stabilization of attractors with different basins of attraction, and also for temporal dynamics of attractor switching that is not possible in standard auto-associative schemes. The dynamics of our models give a plausible account of different sorts of memory retrieval.
\end{abstract}


\maketitle

The brain stores memories as patterns of synaptic strengths in the network of neurons. It can store multiple memories  and retrieve them in a reliable way, and can change from one to another as attention wanders. However, there is no agreement in the neuroscience community of how this occurs. This paper offers a partial solution to understanding the mechanism for retrieval and switching based on a known physiological effect, spike frequency adapation (SFA). 

Decades of work on understanding storage and retrieval have focussed on versions of the Hopfield  model (a special form of the Ising model)  \cite{Amit:1987ws,Hasselmo:1992vn,Hopfield:1982ge}. Hopfield networks have many attractive features: they are  auto-associative: that is, memories are recalled from a fragment of their data because the memories are stored in \emph{attractors}, i.e. metastable states. 

However, as in any statistical model at zero temperature there is no mechanism for escaping an attractor: a single memory pattern would exist for all time. To overcome this problem an artificial 'temperature' is introduced in Hopfield models to allow switching. This `temperature' (i.e fast random noise in synaptic current) has no obvious biological origin. Thus, despite the elegance of the model, and its utility in computer science, its application to the brain is problematic. Previous efforts to over come this limitation have used feedback input \cite{Recanatesi:2015ds}, synaptic depression \cite{Lerner:2014ec}, and adaptive mechanisms \cite{Akrami:2012jg}.
As we will see, SFA allows escape from attractors, and, in some cases acts in the same way as the Hopfield temperature. In addition, it turns off particular memories rather than globally smearing out all of them, as temperature does.

SFA is an activity induced reduction in neural firing rate induced by a hyperpolarizing current that grows as a neuron fires -- neurons that fire a good deal tend to stop firing after a delay. Thus SFA is a natural mechanism to turn off activity. Further, SFA  can be controlled by neuromodulators such as Acetylcholine (ACh), an important regulator of neural excitability. ACh causes a reduction in SFA and provides for its dynamic regulation \cite{Aiken:1995tx,Tang:1997vx}. 
In previous work \cite{Roach:2015dn} we have presented a network model of Hodgkin-Huxley (HH) neurons with SFA and `Mexican Hat' coupling which reproduces many features of cortical activity as ACh levels change between sleep and waking states. In  the present work (below) we use this model to concentrate  on memory retrieval. However, we claim that the essentials of our results are quite robust and independent of the details of the neuron model. To show this, we first consider a version of the  Hopfield model  \cite{Amit:1987ws,Hasselmo:1992vn,Hopfield:1982ge}  which has SFA. 

In our model we consider networks  composed of N=1000 spins, $S=\{s_i\}$ where  $s_{i}=\pm$ 1. The network is  fully connected with weights $\sigma_{i,j}$. As usual, spin up corresponds to a neuron that fires, and spin down to a silent one. Each spin gets an input:  
\begin{equation}
{ h }_{ i }(t)=\sum _{ j=1 }^{ N }{ { \sigma  }_{ i,j }{ s }_{ j } }-\theta_i(t), 
\end{equation}
where $\theta_i(t)$ is a local offset field at site $i$ which changes slowly  in time. The first term is the usual Hopfield-Ising term and the second represents SFA.

The dynamics of the spins are as follows: at each time step a random spin is flipped with probability:
\begin{eqnarray}
P_{h}(s_i)=\frac{1}{1+e^{{- 2 s_i h_{i}}/{T}}},
\end{eqnarray}
where $T$ is the noise. In much of what follows we take $T$ to be very small so that $P_{h}$ is essentially a step function.

\begin{figure}
\includegraphics{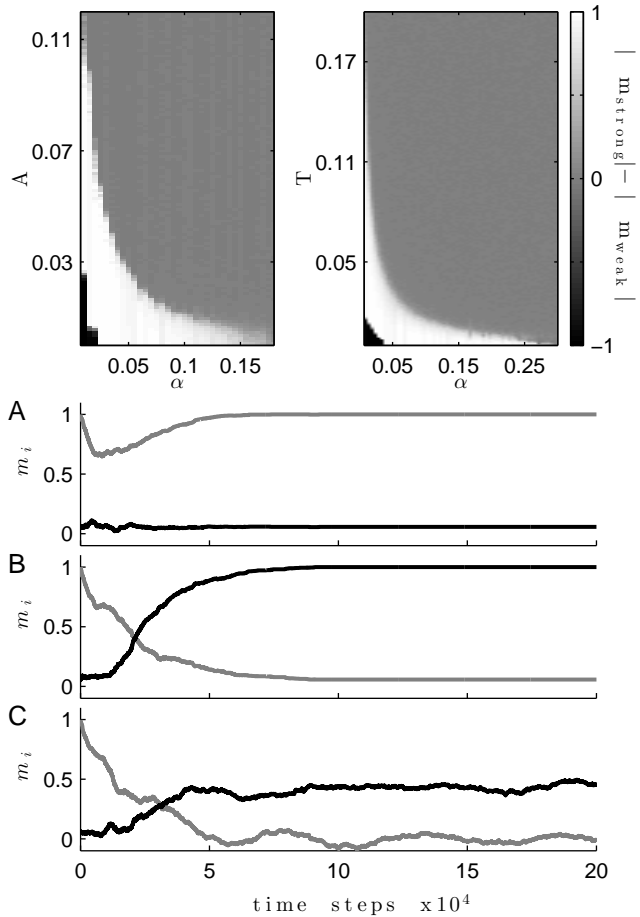}
\caption{\label{fig:hop_phase} SFA and $T$ control attractor stability in Hopfield networks. With noise  Hopfield networks have three  functional states: stability of local attractors, stability of global attractors, and stability of no attractor. The stability of local versus global versus no attractors is shown by the ability of the attractor to move from a weak attractor to a strong one, which is quantified by the difference of the $m_{\mu}$ of the strong attractor and the weak attractor in which the system was initialized. ($Top$)The  saturation  of memories ($\alpha$), noise ($T$), and adaptation ($A$) affect  stability  in a similar manner. For low levels of noise local attractors are stable (black). For a given $\alpha$ either  increasing $A$ (left) or $T$ (right) leads to a strong attractors being stable (white). Further increase destabilizes all attractors (gray). ($Bottom$) Example dynamics of memory overlap, $m_{i}$, for strongly (black) and weakly (gray) weighted memories. In each case $\alpha = 0.01$; adaptation levels are ($A$) A = 0.01 ($B$) A = 0.05 ($C$) A = 0.3.}
\end{figure}

\begin{figure}
\includegraphics{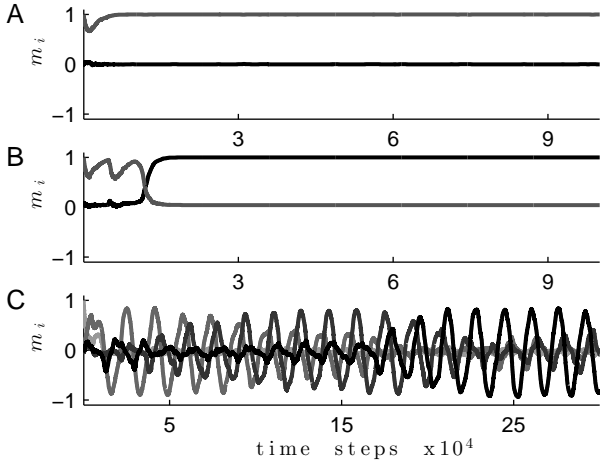}
\caption{\label{fig:ex_dyna} Examples of network dynamics in the modified Hopfield model. For small $A$ even a  weak attractor is stable (panel A; A = 0.01). A moderate increase leads to the strong attractor becoming stable (B, A = 0.1). For larger $A$ damped oscillations with period $\sim 4 \tau_{1}$ emerge: C, A = 0.4. In panels A and B the gray line corresponds to a weak attractor and the black corresponds to a strong one. In panel C, the lightest gray is the weakest attractor. All other lines are strong attractors of equal weight.  }
\end{figure}

The dynamics of  $\theta_i$ is:
\begin{eqnarray}
\theta_{i}(s_i)=\frac{A}{1+e^{{-s_i  (\hat {t}  - \tau_{1})}/{\tau_{2}}}}.
\end{eqnarray}
Here, $\hat{t}$ is the time since the last state change of the spin, and $\tau_{1,2}$ are time constants which govern the dynamics of attractors. The field  $\theta$ increases to $A$ for up spins and decreases to zero for down spins. The time constant $\tau_{1}$  is the time to the half-maximum value of $\theta_{i}$.  We take  $\tau_{1}=5$ (timesteps/N), except for the data in Figure \ref{fig:hop_phase} where $\tau_{1}$ = 1.5 (timesteps/N). The rate at which SFA activates/ deactivates is controlled by $\tau_{2}$ which is set to 0.2, except for the data in Figure \ref{fig:ex_dyna} where $\tau_{2}$ = 0.6 (timesteps/N). This implementations of adaptation is different than others in the hopfield model \cite{Akrami:2012jg}. because it integrates over a longer time (i.e. considers more than the activity at the previous time step). This more closely resembles adaptation in biophysical models.

In the Hopfield scheme memories are stored as attractors, i.e. metastable configurations,  $\Xi^{\mu} = \{\xi^{\mu}_i\}$. We encode attractors using a modified Hebbs rule \cite{Amit:1987ws}:
\begin{eqnarray}
\sigma_{i,j}=\frac{1}{NW}\sum_{\mu}^{p}{w_{\mu}\xi_{i}^{\mu}\xi_{j}^{\mu}}.
\end{eqnarray} 
Each attractor is given a weight, $w_{\mu}$ and $W=\sum_{\mu}^{p}{w_{\mu}}$. Thus $\sigma_{i,j}$ = 1 for two spins with correlated activity across all attractors, $\Xi^{\mu}$, and -1 for spins with anti-correlated activity.  We set $w_{1}/w_{p-1}$ = 0.5 and $w_{p}$ = 1, except for the data reported in  Figure \ref{fig:ex_dyna}D where all  the weights $w_i= 1.0$. The saturation is defined as $\alpha=p/N$. Attractors encoded with lower $w_{\mu}$ are weaker attractors.

In order to determine if the dynamics has settled into the various $\Xi^{\mu}$ we measure the overlap between the stored memory and the current state:
\begin{eqnarray}
{ m }_{\mu} = \frac{1}{N}  \sum _{ i }^{ N }{ { s }_{ i }{ \xi  }_{ i }^{ \mu } } ,  
\end{eqnarray}
which is $\pm1$ when $S = \pm \Xi$  and 0 when $S \bot \Xi$. In each  simulation $S$ was always initialized to a random weak attractor, $\Xi_{weak}$.

In the usual Hopfield model the preference for local, global, or no attractors changes as the noise, $T$, increases \cite{Amit:1987ws,Lewenstein:1989un}. This transition depends on the storage capacity $\alpha$  \cite {Amit:1987ws}. We find  very similar behavior as we increase the magnitude of  SFA, i.e. $A$. To show this 
we compare the standard $T$ versus $\alpha$  plot with  a plot of $A$ versus $\alpha$ in Figure \ref{fig:hop_phase}.  The right panel  shows how $T$ and $\alpha$ interact to affect the stability of the strong and weak attractors and, eventually, to destabilize all attractors.  For small $T$ the dynamics keeps the system in a weak attractor (black on the colormap); for larger $T$  the system enters a  regime of stability of stronger attractors (white). For large $T$ no attractors are stable (gray).

\begin{figure}
\includegraphics{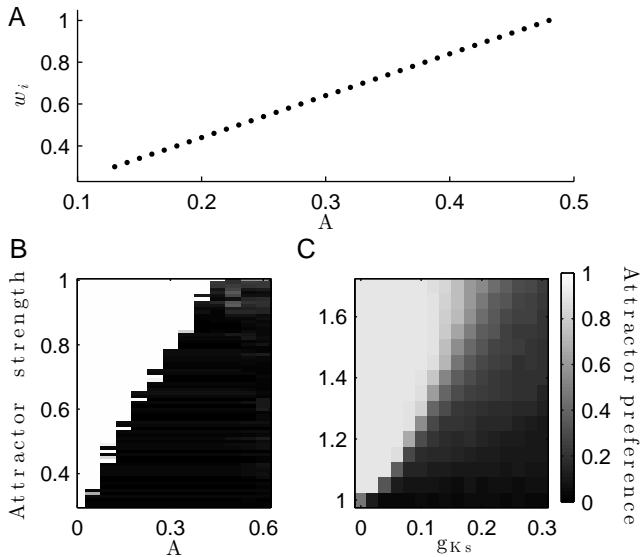}
\caption{\label{fig:leave_fig} Attractor stability varies as a function of $A$. For SFA to induce the network to leave a attractor it must be large enough to overcome the energy barrier of the attractor. Mean field calculations predict a linear relationship between the strength of an attractor and the amount of adaptation, $A$, to destabilize it (A).This is best seen in the Hopfield model (B). The threshold value of $A$ increases linearly as the attractor strength, $w_{\mu}$ increases. A similar effect is seen in the spiking network model (C ). }
\end{figure}

\begin{figure}
\includegraphics{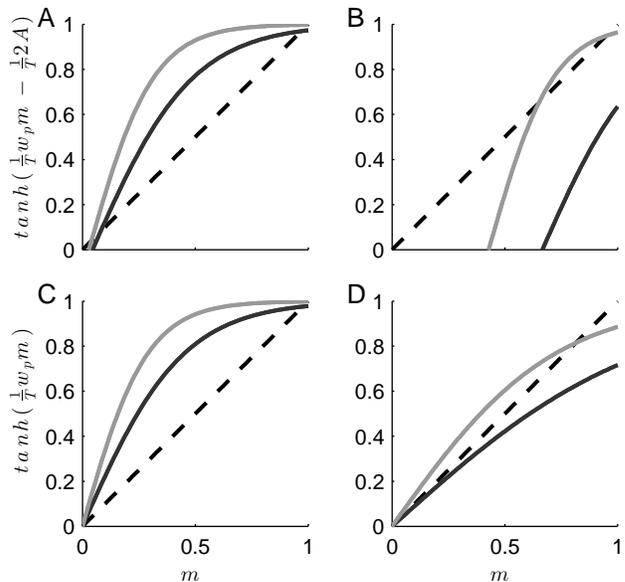}
\caption{\label{fig:ana_fig} SFA and $T$ destabilize attractors of different strengths in Hopfield networks. Solutions to mean field equations illustrate how adaptation and temperature destabilize weak attractors. When $A$ is low many memories are stabile (panel A; A=0.1). Increasing $A$ destabilizes weak memories while preserving strong memories (panel B; A=0.25). When adaptation is absent temperature has a similar effect where all memories are stable for low $T$ (panel C; $T=0.2$), while only strong memories are stable for high $T$ (panel D; $T=0.5$). The dashed line shows $m=m$; the solid black line shows the mean field equation for $w_{\upsilon}=0.45$; the solid gray line shows the mean field equation for $w_{\upsilon}=0.75$.}
\end{figure}

SFA has a very similar effect; see Figure \ref{fig:hop_phase}, left.  To see how this comes about, consider the Hamiltonian:
\begin{eqnarray}
E=-\frac{1}{2}\sum_{i,j}^{N}{\sigma_{i,j}s_{i}s_{j}}+\sum_{i}^{N}{\theta_{i}(t)s_{i}},
\end{eqnarray}
which defines a slowly varying energy landscape. The first term is the ordinary Ising energy, and the second can be thought of as a time-dependent magnetic field that increases (for up spins) to $A$ when $t \gg \tau_2$. Increasing $A$  destabilizes minima in $E$; an  attractor becomes unstable for $t\rightarrow\infty$ if $A$ is large enough. In Figure \ref{fig:leave_fig} we show the magnitude of SFA required to cause the system to leave an attractor of a given strength. It increases linearly as attractor strength increases. A similar effect occurs in the HH model, below. 

The stability of an attractor of a given weight can also be investigated by mean field theory \cite{Amit:1985um,Lewenstein:1989un}. The mean field equations for the system are:

\begin{eqnarray}
\left< s_{i} \right> = \tanh(\frac{\beta}{NW}\sum_{j,\mu}{w_{\mu}\xi_{i}^{\mu}\xi_{j}^{\mu}\left< s_{j} \right>} -2\theta_{i}),
\end{eqnarray}
where $\beta = \sfrac{1}{T}$. By exploiting the fact that $\left< s_{i} \right> = m\xi^{\upsilon}_{i}$ the mean field equations can be rewritten as:
\begin{eqnarray}
m\xi^{\upsilon}_{i} = \tanh(\frac{\beta}{NW}\sum_{j,\mu}{w_{\mu}\xi_{i}^{\mu}\xi_{j}^{\mu}m\xi^{\upsilon}_{i} - 2\theta_{i}}).
\end{eqnarray}
If $p \ll N$ any overlap between memories is negligible so the mean field equation and the system is in memory $\upsilon$ for a time $ \gg \tau_{1}$ becomes:
\begin{eqnarray}
m\xi^{\upsilon}_{i}=\tanh(\beta w_{\upsilon}m\xi^{\upsilon}_{i} - \frac{1}{T}2A), 
\end{eqnarray}
which be simplified to $m=\tanh(\beta w_{\upsilon}m -\beta2A )$. When this equation has solutions beyond $m=0$ a memory with strength $w_{\upsilon}$ is stable for a given $T$ or $A$. Figure \ref{fig:ana_fig} shows mean field solutions for memories with strengths $w_{\upsilon} = 0.45$ (solid black line) and $w_{\upsilon} = 0.75$ (solid gray line). As in the numerical results adaptation (\ref{fig:ana_fig} top panels) and temperature (bottom panels) have similar effects on the stability of memories. For low levels ($A=0.1$, $T=0.2$; left panels) both strong and weak memories are stable (i.e. both have solutions beyond $m=0$), but moderate increases in $A$ or $T$ destabilize weaker memories ($A=.25$, $T=0.5$; right panels)

Thus, changes in the strength of SFA can play the same role as changes in $T$ by  destabilizing attractors of varying strength as $A$ increases. Interesting time-dependent effects occur for  
intermediate values of $A$ when the $\tau$'s are not too large. Because $\theta_i$ is a function of $t$ we can generate chains of attractor preferences, as opposed to stability in a deep attractor or a random walk (as in the standard  Hopfield model for large $T$). These results are shown in Figure \ref{fig:ex_dyna}. For small $A$ local, weak attractors are stable.  A moderate increase leads to  strong attractors being stable (Figure \ref{fig:ex_dyna} A, B). Further increase of $A$ leads to oscillations of period $\sim 4\tau_{1}$; Figure \ref{fig:ex_dyna} C. This is similar to the latching dynamics found in \cite{Akrami:2012jg}.  

We next turn to a more realistic neuron model to compare the effects of SFA for the two cases. The spiking network model introduced previously \cite{Roach:2015dn} considers  $N_{E}=1225$ excitatory and $N_{I}=324$ inhibitory HH neurons arrayed on two square lattices of size $L_{E/I}$. The coupling was of lateral inhibition (Mexican Hat) type where short range excitation is balanced with global inhibition. All excitatory neurons were connected to neighbors within radius $R_{xx}=\sqrt { { { L }_{ E/I }^{ 2 }{ k }_{ xx } }/{ \pi { N }_{ E/I } } } $ where $k_{ei}=16$ is the degree of excitatory to excitatory connections, $k_{ei}=4$ is the degree of excitatory to inhibitory connections. Neural dynamics were modeled by the current balance equation \cite{Stiefel:2008dv}:

\begin{eqnarray}
{ c }_{ m }\frac { dV_{i} }{ dt } =-{ g }_{ Na }{ m }_{ \infty  }^{ 3 }h(V_{i}-{ E }_{ Na })-{ g }_{ Kdir }{ n }^{ 4 }(V_{i}-{ E }_{ K })\nonumber\\
-{ g }_{ Ks }s(V_{i}-{ E }_{ K })-g_{L}(V_{i}-E_{L})-{ I }_{ syn,i }+{ I }_{ ext } \quad
\end{eqnarray}
In this equation, as we will see, $g_{Ks}$ sets the magnitude of the SFA; it corresponds to $A$ in the model above.

The dynamics of the gating variables $h$, $n$ and $s$ is of the form $dx/dt = (x_{\infty}(V) - x)/\tau_{x}(V)$ with additional specific evolution of the two voltage dependent parameters $x_{\infty}$ and $\tau_{x}$.  The slow potassium current conductance, $g_{Ks}$ controls the level of SFA (i.e. lower values of $g_{Ks}$ correspond to low SFA). The level of ACh modulates $g_{Ks}$:  the maximum ($g_{Ks}=1.5$ mS/cm$^2$)  and minimum ($g_{Ks}=0$) correspond to the absence or maximum of ACh, respectively. For more details see \cite{Roach:2015dn}.

The synaptic current to neuron $i$ is ${ I }_{ syn,i } = g_{E}(t)(V_{i}-E_{E}) + g_{I}(t)(V_{i}-E_{I})$ and the dynamics of $g_{E/I}(t)$ is:
\begin{eqnarray}
{ g }_{ E/I }(t)=K\sum _{ j }^{ \in E/I }{ { \sigma  }_{ i,j }({ e }^{ \frac { -({ \tilde { t }  }_{ j }-{ \tau  }_{ D }) }{ { \tau  }_{ S } }  }-{ e }^{ \frac { -({ \tilde { t }  }_{ j }-{ \tau  }_{ D }) }{ { \tau  }_{ F } }  }) } 
\end{eqnarray}
where $\sigma_{i,j}$ is the synaptic weight between neuron $i$ and neuron $j$, and $\tau_{S,F}$ are time constants equal to 3.0 and 0.3 ms respectively. $\sigma_{i,j}$ is set to 0.02 $\rm mS/cm^2$ unless otherwise stated, $\tilde{t}_{j}$ is the time since the last spike of neuron $j$, and $K$ is a normalization constant. $I_{ext}$  set so that all neurons  fire at 10 Hz in the absence of synaptic input and $g_{Ks}$. The equations were  integrated using the 4th order Runge-Kutta method at a 0.05 ms time step to 20 s. Data points are averages of 20 sets of initial conditions.

We have shown \cite{Roach:2015dn} that the nature of the dynamics is that for small  $g_{Ks}$ (large concentrations of ACh) there is a stationary, localized region (a `bump') of spiking activity. For larger $g_{Ks}$ the bump travels through the lattice. 

\begin{figure}
\includegraphics{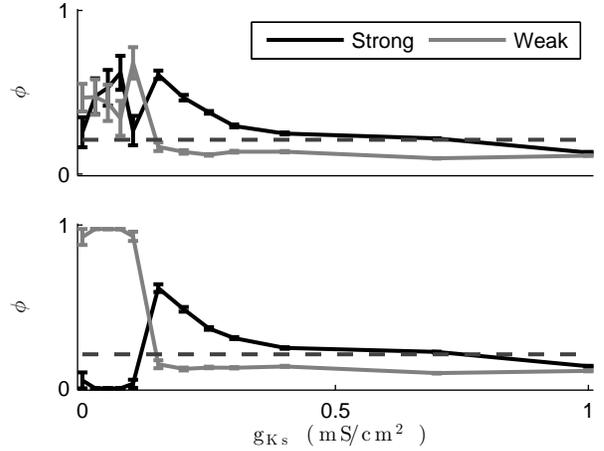}
\caption{ \label{fig:phi_pref} Attractor preference and $g_{Ks}$. The quantity $\phi$ is the fraction of time that activity is located within an attractor. Black dashed line, control value. (Top) For initial locations   outside any attractor no clear preference emerges for small $g_{Ks}$. For moderate  $g_{Ks}$  there is clear preference for the strong attractor.  (Bottom) For initial conditions   within the weak attractor activity never leaves  for small $g_{Ks}$. There is  significant preference for the strong attractor for moderate $g_{Ks}$.} 
\end{figure}

To consider memory we introduce spatial attractors by increasing synaptic strength in certain locations \cite{Renart:2003tu,Roach:2015dn}. These attractors fix the location of the bump when the dynamics is in the stationary regime. For a single attractor,  preference for the attractor  falls as $g_{Ks}$ increases \cite{Roach:2015dn}. This model is quite different from the one discussed above: the excitatory coupling is short-ranged in contrast to the Hopfield $\sigma$'s which are  long-ranged. The attractors here are defined by local geometry. Nevertheless, SFA gives common results for the two cases. 

To consider multiple attractors of variable strength we strengthened synaptic strength in two network regions at opposite ends of the lattice. The strong attractor had 100\% stronger excitatory connections and the weak attractor had a 50\% increase. 
To examine how network preference changed as a function of $g_{Ks}$ in multi-attractor networks we did  simulations where activity was initialized by injecting a 0.25 $\rm \mu A/cm^{2}$ current to a region outside either of the attractors for the first 0.5 s of the simulation. Preference for a given attractor was quantified by the measure $\phi$ which is the fraction of time that the center of the bump, calculated according to \cite{Bai:2008ex}, is located within the attractor. For low levels of SFA  there is no clear preference indicating that activity localizes to attractors randomly; Figure \ref{fig:phi_pref}, top. Increasing $g_{Ks}$ leads to a clear preference for the strong attractor. The preference for any attractor disappears for large $g_{Ks}$ as the network enters the regime of traveling bump dynamics.

To further test the stability of the two attractors we put activity initially on the weak attractor. For small $g_{Ks}$  activity remained localized there: Figure \ref{fig:phi_pref}, bottom.  For larger  $g_{Ks}$ activity moved to the stronger attractor. This confirms that the strength of SFA can control the stability of attractors having different depths. 

\begin{figure}
\includegraphics{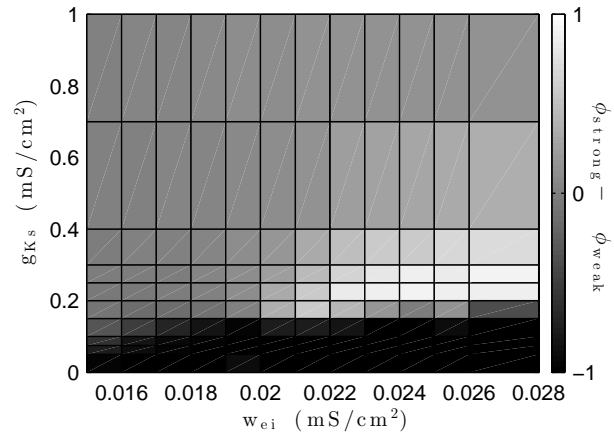}
\caption{\label{fig:phi_inh} Strong attractor preference is controlled by inhibition strength in spiking networks. We measure the differential attractor preference for the two attractors by  ($\phi_{strong} \quad - \quad \phi_{weak}$), which is 1 when all activity is located within the strong attractor and -1 when all activity is within the weak. The preference for the strong attractor at moderate SFA disappears when inhibition is decreased.}
\end{figure}

We also considered how the relative stability of the attractors depends on the ratio of inhibitory to excitatory coupling, $w_{e/i}$).  Figure \ref{fig:phi_inh} is a phase plot of final attractor preference for different $w_{e/i}$) and $g_{Ks}$. The relative preference for the two attractors was measured by  $\phi_{strong} - \phi_{weak}$, which ranges between 1 (preference for the strong attractor) and -1 (preference for the weak). Interestingly, weakening inhibition abolishes any preference for the strong attractor at intermediate levels of $g_{Ks}$.   

Thus, for both models SFA can selectively destabilize attractors effectively controlling attractor preference.  With SFA we can have long-term preferential activation of attractors of different strengths and also  non-trivial time dependence of attractor sequences. This provides a  biologically plausible mechanism for switching between encoded patterns \cite{Lewenstein:1989un,Monasson:2015kc}.

In the biophysical model SFA depends on ACh \cite{Stiefel:2008dv,Tang:1997vx,Tsuno:2013kh}.  Our results imply that ACh controls memory retrieval dynamics. Note the relevance of our results to context dependent release of ACh and its role in attention \cite{Hasselmo:2011dt}. During tasks requiring a high degree of focus, low SFA allows the brain to fix on the memory that closely fits the current sensory input. On the other hand, with high ACh  the attractor can be reinforced by synaptic plasticity. As attention requirements are relaxed, and ACh levels fall, moderate levels of SFA allow for sampling of the memory space, see \cite{Hopfield:2010kf}. 

The largest variation in cortical ACh levels occurs between sleep/ wake states. In this case  the highest levels occur during rapid eye movement (REM) sleep and the lowest during slow wave sleep \cite{Vazquez:2001tg}. We argue that intermediate levels of ACh during wake states allow for memory recall when externally driven network states are allowed to wander to find the optimal state. REM sleep is thought to be important for memory consolidation, where retrieval of weakly stored attractors of previous experience is essential to their consolidation.  NREM sleep associated with low ACh levels is characterized by slow waves and may play a role in synaptic rescaling \cite{Tononi:2003hh}.

\begin{acknowledgments}

JPR was supported by an NSF Graduate Research Fellowship Program under Grant No. DGE 1256260 and a UM Rackham Merit Fellowship. MRZ and LMS were supported by NSF PoLS 1058034.
\end{acknowledgments}


\begin{thebibliography}{20}%
\makeatletter
\providecommand \@ifxundefined [1]{%
 \@ifx{#1\undefined}
}%
\providecommand \@ifnum [1]{%
 \ifnum #1\expandafter \@firstoftwo
 \else \expandafter \@secondoftwo
 \fi
}%
\providecommand \@ifx [1]{%
 \ifx #1\expandafter \@firstoftwo
 \else \expandafter \@secondoftwo
 \fi
}%
\providecommand \natexlab [1]{#1}%
\providecommand \enquote  [1]{``#1''}%
\providecommand \bibnamefont  [1]{#1}%
\providecommand \bibfnamefont [1]{#1}%
\providecommand \citenamefont [1]{#1}%
\providecommand \href@noop [0]{\@secondoftwo}%
\providecommand \href [0]{\begingroup \@sanitize@url \@href}%
\providecommand \@href[1]{\@@startlink{#1}\@@href}%
\providecommand \@@href[1]{\endgroup#1\@@endlink}%
\providecommand \@sanitize@url [0]{\catcode `\\12\catcode `\$12\catcode
  `\&12\catcode `\#12\catcode `\^12\catcode `\_12\catcode `\%12\relax}%
\providecommand \@@startlink[1]{}%
\providecommand \@@endlink[0]{}%
\providecommand \url  [0]{\begingroup\@sanitize@url \@url }%
\providecommand \@url [1]{\endgroup\@href {#1}{\urlprefix }}%
\providecommand \urlprefix  [0]{URL }%
\providecommand \Eprint [0]{\href }%
\providecommand \doibase [0]{http://dx.doi.org/}%
\providecommand \selectlanguage [0]{\@gobble}%
\providecommand \bibinfo  [0]{\@secondoftwo}%
\providecommand \bibfield  [0]{\@secondoftwo}%
\providecommand \translation [1]{[#1]}%
\providecommand \BibitemOpen [0]{}%
\providecommand \bibitemStop [0]{}%
\providecommand \bibitemNoStop [0]{.\EOS\space}%
\providecommand \EOS [0]{\spacefactor3000\relax}%
\providecommand \BibitemShut  [1]{\csname bibitem#1\endcsname}%
\let\auto@bib@innerbib\@empty
\bibitem [{\citenamefont {Amit}\ \emph {et~al.}(1987)\citenamefont {Amit},
  \citenamefont {Gutfreund},\ and\ \citenamefont {Sompolinsky}}]{Amit:1987ws}%
  \BibitemOpen
  \bibfield  {author} {\bibinfo {author} {\bibfnamefont {D.~J.}\ \bibnamefont
  {Amit}}, \bibinfo {author} {\bibfnamefont {H.}~\bibnamefont {Gutfreund}}, \
  and\ \bibinfo {author} {\bibfnamefont {H.}~\bibnamefont {Sompolinsky}},\
  }\href@noop {} {\bibfield  {journal} {\bibinfo  {journal} {Annals of
  Physics}\ }\textbf {\bibinfo {volume} {173}},\ \bibinfo {pages} {30}
  (\bibinfo {year} {1987})}\BibitemShut {NoStop}%
\bibitem [{\citenamefont {Hasselmo}\ \emph {et~al.}(1992)\citenamefont
  {Hasselmo}, \citenamefont {Anderson},\ and\ \citenamefont
  {Bower}}]{Hasselmo:1992vn}%
  \BibitemOpen
  \bibfield  {author} {\bibinfo {author} {\bibfnamefont {M.~E.}\ \bibnamefont
  {Hasselmo}}, \bibinfo {author} {\bibfnamefont {B.~P.}\ \bibnamefont
  {Anderson}}, \ and\ \bibinfo {author} {\bibfnamefont {J.~M.}\ \bibnamefont
  {Bower}},\ }\href@noop {} {\bibfield  {journal} {\bibinfo  {journal} {Journal
  of Neurophysiology}\ }\textbf {\bibinfo {volume} {67}},\ \bibinfo {pages}
  {1230} (\bibinfo {year} {1992})}\BibitemShut {NoStop}%
\bibitem [{\citenamefont {Hopfield}(1982)}]{Hopfield:1982ge}%
  \BibitemOpen
  \bibfield  {author} {\bibinfo {author} {\bibfnamefont {J.~J.}\ \bibnamefont
  {Hopfield}},\ }\href@noop {} {\bibfield  {journal} {\bibinfo  {journal}
  {Proceedings of the National Academy of Sciences}\ }\textbf {\bibinfo
  {volume} {79}},\ \bibinfo {pages} {2554} (\bibinfo {year}
  {1982})}\BibitemShut {NoStop}%
\bibitem [{\citenamefont {Recanatesi}\ \emph {et~al.}(2015)\citenamefont
  {Recanatesi}, \citenamefont {Katkov}, \citenamefont {Romani},\ and\
  \citenamefont {Tsodyks}}]{Recanatesi:2015ds}%
  \BibitemOpen
  \bibfield  {author} {\bibinfo {author} {\bibfnamefont {S.}~\bibnamefont
  {Recanatesi}}, \bibinfo {author} {\bibfnamefont {M.}~\bibnamefont {Katkov}},
  \bibinfo {author} {\bibfnamefont {S.}~\bibnamefont {Romani}}, \ and\ \bibinfo
  {author} {\bibfnamefont {M.}~\bibnamefont {Tsodyks}},\ }\href@noop {}
  {\bibfield  {journal} {\bibinfo  {journal} {Frontiers in computational
  neuroscience}\ }\textbf {\bibinfo {volume} {9}},\ \bibinfo {pages} {275}
  (\bibinfo {year} {2015})}\BibitemShut {NoStop}%
\bibitem [{\citenamefont {Lerner}\ and\ \citenamefont
  {Shriki}(2014)}]{Lerner:2014ec}%
  \BibitemOpen
  \bibfield  {author} {\bibinfo {author} {\bibfnamefont {I.}~\bibnamefont
  {Lerner}}\ and\ \bibinfo {author} {\bibfnamefont {O.}~\bibnamefont
  {Shriki}},\ }\href@noop {} {\bibfield  {journal} {\bibinfo  {journal}
  {Frontiers in psychology}\ }\textbf {\bibinfo {volume} {5}},\ \bibinfo
  {pages} {314} (\bibinfo {year} {2014})}\BibitemShut {NoStop}%
\bibitem [{\citenamefont {Akrami}\ \emph {et~al.}(2012)\citenamefont {Akrami},
  \citenamefont {Russo},\ and\ \citenamefont {Treves}}]{Akrami:2012jg}%
  \BibitemOpen
  \bibfield  {author} {\bibinfo {author} {\bibfnamefont {A.}~\bibnamefont
  {Akrami}}, \bibinfo {author} {\bibfnamefont {E.}~\bibnamefont {Russo}}, \
  and\ \bibinfo {author} {\bibfnamefont {A.}~\bibnamefont {Treves}},\
  }\href@noop {} {\bibfield  {journal} {\bibinfo  {journal} {Brain research}\
  }\textbf {\bibinfo {volume} {1434}},\ \bibinfo {pages} {4} (\bibinfo {year}
  {2012})}\BibitemShut {NoStop}%
\bibitem [{\citenamefont {Aiken}\ \emph {et~al.}(1995)\citenamefont {Aiken},
  \citenamefont {Lampe}, \citenamefont {Murphy},\ and\ \citenamefont
  {Brown}}]{Aiken:1995tx}%
  \BibitemOpen
  \bibfield  {author} {\bibinfo {author} {\bibfnamefont {S.~P.}\ \bibnamefont
  {Aiken}}, \bibinfo {author} {\bibfnamefont {B.~J.}\ \bibnamefont {Lampe}},
  \bibinfo {author} {\bibfnamefont {P.~A.}\ \bibnamefont {Murphy}}, \ and\
  \bibinfo {author} {\bibfnamefont {B.~S.}\ \bibnamefont {Brown}},\ }\href@noop
  {} {\bibfield  {journal} {\bibinfo  {journal} {British journal of
  pharmacology}\ }\textbf {\bibinfo {volume} {115}},\ \bibinfo {pages} {1163}
  (\bibinfo {year} {1995})}\BibitemShut {NoStop}%
\bibitem [{\citenamefont {Tang}\ \emph {et~al.}(1997)\citenamefont {Tang},
  \citenamefont {Bartels},\ and\ \citenamefont {Sejnowski}}]{Tang:1997vx}%
  \BibitemOpen
  \bibfield  {author} {\bibinfo {author} {\bibfnamefont {A.~C.}\ \bibnamefont
  {Tang}}, \bibinfo {author} {\bibfnamefont {A.~M.}\ \bibnamefont {Bartels}}, \
  and\ \bibinfo {author} {\bibfnamefont {T.~J.}\ \bibnamefont {Sejnowski}},\
  }\href@noop {} {\bibfield  {journal} {\bibinfo  {journal} {Cerebral Cortex}\
  }\textbf {\bibinfo {volume} {7}},\ \bibinfo {pages} {502} (\bibinfo {year}
  {1997})}\BibitemShut {NoStop}%
\bibitem [{\citenamefont {Roach}\ \emph {et~al.}(2015)\citenamefont {Roach},
  \citenamefont {Ben-Jacob}, \citenamefont {Sander},\ and\ \citenamefont
  {Zochowski}}]{Roach:2015dn}%
  \BibitemOpen
  \bibfield  {author} {\bibinfo {author} {\bibfnamefont {J.~P.}\ \bibnamefont
  {Roach}}, \bibinfo {author} {\bibfnamefont {E.}~\bibnamefont {Ben-Jacob}},
  \bibinfo {author} {\bibfnamefont {L.~M.}\ \bibnamefont {Sander}}, \ and\
  \bibinfo {author} {\bibfnamefont {M.}~\bibnamefont {Zochowski}},\ }\href@noop
  {} {\bibfield  {journal} {\bibinfo  {journal} {PLoS Computational Biology}\
  }\textbf {\bibinfo {volume} {11}},\ \bibinfo {pages} {e1004449} (\bibinfo
  {year} {2015})}\BibitemShut {NoStop}%
\bibitem [{\citenamefont {Lewenstein}\ and\ \citenamefont
  {Nowak}(1989)}]{Lewenstein:1989un}%
  \BibitemOpen
  \bibfield  {author} {\bibinfo {author} {\bibfnamefont {M.}~\bibnamefont
  {Lewenstein}}\ and\ \bibinfo {author} {\bibfnamefont {A.}~\bibnamefont
  {Nowak}},\ }\href@noop {} {\bibfield  {journal} {\bibinfo  {journal}
  {Physical Review Letters}\ }\textbf {\bibinfo {volume} {62}},\ \bibinfo
  {pages} {225} (\bibinfo {year} {1989})}\BibitemShut {NoStop}%
\bibitem [{\citenamefont {Amit}\ \emph {et~al.}(1985)\citenamefont {Amit},
  \citenamefont {Gutfreund},\ and\ \citenamefont {Sompolinsky}}]{Amit:1985um}%
  \BibitemOpen
  \bibfield  {author} {\bibinfo {author} {\bibfnamefont {D.J.}~\bibnamefont
  {Amit}}, \bibinfo {author} {\bibfnamefont {H.}~\bibnamefont {Gutfreund}}, \
  and\ \bibinfo {author} {\bibfnamefont {H.}~\bibnamefont {Sompolinsky}},\
  }\href@noop {} {\bibfield  {journal} {\bibinfo  {journal} {Physical review.
  A}\ }\textbf {\bibinfo {volume} {32}},\ \bibinfo {pages} {1007} (\bibinfo
  {year} {1985})}\BibitemShut {NoStop}%
\bibitem [{\citenamefont {Stiefel}\ \emph {et~al.}(2008)\citenamefont
  {Stiefel}, \citenamefont {Gutkin},\ and\ \citenamefont
  {Sejnowski}}]{Stiefel:2008dv}%
  \BibitemOpen
  \bibfield  {author} {\bibinfo {author} {\bibfnamefont {K.~M.}\ \bibnamefont
  {Stiefel}}, \bibinfo {author} {\bibfnamefont {B.~S.}\ \bibnamefont {Gutkin}},
  \ and\ \bibinfo {author} {\bibfnamefont {T.~J.}\ \bibnamefont {Sejnowski}},\
  }\href@noop {} {\bibfield  {journal} {\bibinfo  {journal} {Journal of
  Computational Neuroscience}\ }\textbf {\bibinfo {volume} {26}},\ \bibinfo
  {pages} {289} (\bibinfo {year} {2008})}\BibitemShut {NoStop}%
\bibitem [{\citenamefont {Renart}\ \emph {et~al.}(2003)\citenamefont {Renart},
  \citenamefont {Song},\ and\ \citenamefont {Wang}}]{Renart:2003tu}%
  \BibitemOpen
  \bibfield  {author} {\bibinfo {author} {\bibfnamefont {A.}~\bibnamefont
  {Renart}}, \bibinfo {author} {\bibfnamefont {P.}~\bibnamefont {Song}}, \ and\
  \bibinfo {author} {\bibfnamefont {X.-J.}\ \bibnamefont {Wang}},\ }\href@noop
  {} {\bibfield  {journal} {\bibinfo  {journal} {Neuron}\ }\textbf {\bibinfo
  {volume} {38}},\ \bibinfo {pages} {473} (\bibinfo {year} {2003})}\BibitemShut
  {NoStop}%
\bibitem [{\citenamefont {Bai}\ and\ \citenamefont {Breen}(2008)}]{Bai:2008ex}%
  \BibitemOpen
  \bibfield  {author} {\bibinfo {author} {\bibfnamefont {L.}~\bibnamefont
  {Bai}}\ and\ \bibinfo {author} {\bibfnamefont {D.}~\bibnamefont {Breen}},\
  }\href@noop {} {\bibfield  {journal} {\bibinfo  {journal} {Journal of
  Graphics, GPU, and Game Tools}\ }\textbf {\bibinfo {volume} {13}},\ \bibinfo
  {pages} {53} (\bibinfo {year} {2008})}\BibitemShut {NoStop}%
\bibitem [{\citenamefont {Monasson}\ and\ \citenamefont
  {Rosay}(2015)}]{Monasson:2015kc}%
  \BibitemOpen
  \bibfield  {author} {\bibinfo {author} {\bibfnamefont {R.}~\bibnamefont
  {Monasson}}\ and\ \bibinfo {author} {\bibfnamefont {S.}~\bibnamefont
  {Rosay}},\ }\href@noop {} {\bibfield  {journal} {\bibinfo  {journal}
  {Physical Review Letters}\ }\textbf {\bibinfo {volume} {115}},\ \bibinfo
  {pages} {098101} (\bibinfo {year} {2015})}\BibitemShut {NoStop}%
\bibitem [{\citenamefont {Tsuno}\ \emph {et~al.}(2013)\citenamefont {Tsuno},
  \citenamefont {Schultheiss},\ and\ \citenamefont {Hasselmo}}]{Tsuno:2013kh}%
  \BibitemOpen
  \bibfield  {author} {\bibinfo {author} {\bibfnamefont {Y.}~\bibnamefont
  {Tsuno}}, \bibinfo {author} {\bibfnamefont {N.~W.}\ \bibnamefont
  {Schultheiss}}, \ and\ \bibinfo {author} {\bibfnamefont {M.~E.}\ \bibnamefont
  {Hasselmo}},\ }\href@noop {} {\bibfield  {journal} {\bibinfo  {journal} {The
  Journal of Physiology}\ }\textbf {\bibinfo {volume} {591}},\ \bibinfo {pages}
  {2611} (\bibinfo {year} {2013})}\BibitemShut {NoStop}%
\bibitem [{\citenamefont {Hasselmo}\ and\ \citenamefont
  {Sarter}(2011)}]{Hasselmo:2011dt}%
  \BibitemOpen
  \bibfield  {author} {\bibinfo {author} {\bibfnamefont {M.~E.}\ \bibnamefont
  {Hasselmo}}\ and\ \bibinfo {author} {\bibfnamefont {M.}~\bibnamefont
  {Sarter}},\ }\href@noop {} {\bibfield  {journal} {\bibinfo  {journal}
  {Neuropsychopharmacology}\ }\textbf {\bibinfo {volume} {36}},\ \bibinfo
  {pages} {52} (\bibinfo {year} {2011})}\BibitemShut {NoStop}%
\bibitem [{\citenamefont {Hopfield}(2010)}]{Hopfield:2010kf}%
  \BibitemOpen
  \bibfield  {author} {\bibinfo {author} {\bibfnamefont {J.~J.}\ \bibnamefont
  {Hopfield}},\ }\href@noop {} {\bibfield  {journal} {\bibinfo  {journal}
  {Proceedings of the National Academy of Sciences}\ }\textbf {\bibinfo
  {volume} {107}},\ \bibinfo {pages} {1648} (\bibinfo {year}
  {2010})}\BibitemShut {NoStop}%
\bibitem [{\citenamefont {Vazquez}\ and\ \citenamefont
  {Baghdoyan}(2001)}]{Vazquez:2001tg}%
  \BibitemOpen
  \bibfield  {author} {\bibinfo {author} {\bibfnamefont {J.}~\bibnamefont
  {Vazquez}}\ and\ \bibinfo {author} {\bibfnamefont {H.~A.}\ \bibnamefont
  {Baghdoyan}},\ }\href@noop {} {\bibfield  {journal} {\bibinfo  {journal}
  {American journal of physiology. Regulatory, integrative and comparative
  physiology}\ }\textbf {\bibinfo {volume} {280}},\ \bibinfo {pages} {R598}
  (\bibinfo {year} {2001})}\BibitemShut {NoStop}%
\bibitem [{\citenamefont {Tononi}\ and\ \citenamefont
  {Cirelli}(2003)}]{Tononi:2003hh}%
  \BibitemOpen
  \bibfield  {author} {\bibinfo {author} {\bibfnamefont {G.}~\bibnamefont
  {Tononi}}\ and\ \bibinfo {author} {\bibfnamefont {C.}~\bibnamefont
  {Cirelli}},\ }\href@noop {} {\bibfield  {journal} {\bibinfo  {journal} {Brain
  Research Bulletin}\ }\textbf {\bibinfo {volume} {62}},\ \bibinfo {pages}
  {143} (\bibinfo {year} {2003})}\BibitemShut {NoStop}%
\end{thebibliography}
%

\end{document}